\documentstyle[twoside,fleqn,espcrc2,epsfig]{article}

\title{Ultra-Peripheral Collisions with STAR at RHIC}

\author{Spencer Klein\address[LBNL]{Lawrence Berkeley National
Laboratory, Berkeley, CA 94720}, for the STAR collaboration}

\begin{document}

\begin{abstract}

The strong electromagnetic fields of heavy nuclei can produce a wide
variety of two-photon and photonuclear reactions at relativistic ion
colliders.  We present recent results from the STAR collaboration on
these `ultra-peripheral' interactions, focusing on vector meson
production and interferometry, and on $e^+e^-$ pair production.  The
vector meson interferometry occurs because of the symmetric initial
state: nucleus 1 can emit a photon which scatters from nucleus 2,
emerging as a vector meson, or vice-versa.  The two processes are
indistinguishable, and so interfere, even though the production points
are separated enough that the produced mesons decay before their wave
functions can overlap, so the system can be used for interesting tests
of quantum mechanics.

\vspace{1pc}
\end{abstract}

\maketitle
\section{Introduction}

Electromagnetic interactions between relativistic heavy ions occur
frequently, even at impact parameters $b$ large enough that no
hadronic interactions occur \cite{reviews}.  The electromagnetic
fields may be treated as fields of almost-real photons, following the
Weizs\"acker-Williams method; the photon flux scales as $Z^2$, so the
cross sections are large.

These ultra-peripheral collisions (UPCs) can be divided into two
classes, purely electromagnetic interactions (two-photon) and
photonuclear interactions, where a photon from the field of one
nucleus interacts with the other nucleus.  We focus on the region
$b>2R_A$ to exclude hadronic contamination. 
We will present 3 topics which have been studied by the
Solenoidal Tracker at RHIC (STAR) collaboration: vector meson
production, vector meson interferometry, and $e^+e^-$ pair production.

\begin{figure}
\label{diagram}
\setlength{\epsfxsize=3.0 in} 
\centerline{\epsffile{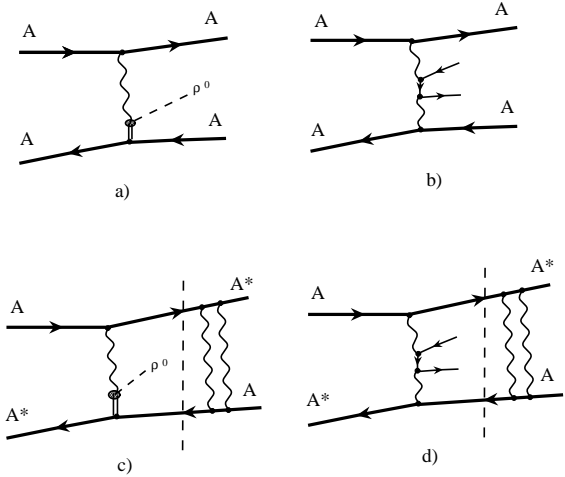}}
\caption[]{Schematic diagrams for (a) $\rho^0$ production and (b)
$e^+e^-$ production.  These processes may be accompanied by mutual
Coulomb excitation, as in (c) and (d) respectively.  The dashed lines
in (c) and (d) show the factorization, Eq.  \ref{eq:factorize}.}
\end{figure}

Vector mesons are produced when a photon from the field of one nucleus
fluctuates to a $q\overline q$ pair (virtual vector meson), which then
scatters elastically from the other nucleus, emerging as a vector
meson \cite{phase}, as is shown in Fig. \ref{diagram}a.  The elastic
scattering is colorless, and can be described in terms of Pomeron
exchange.  For light mesons like the $\rho^0$, the cross section rises
slowly with the photon energy $k$.  For small momentum transfers, the
elastic scattering is coherent over the entire nuclear target.  In the
coherent region, the cross section depends on the size of the
$q\overline q$ dipole fluctuation.  

Heavy mesons come from small dipoles with small interaction cross
sections.  Production is distributed evenly throughout the target, and
$\sigma \approx A^2$.  For sufficiently heavy mesons like the
$J/\psi$, the elastic scattering may be described in terms of multiple
gluon exchange.  The cross section is sensitive to gluon shadowing in
nuclei \cite{strikman}.

Large dipoles (light mesons) interact with a large cross section with
the first nucleon that they encounter.  The interaction is on the
surface of the target, so $\sigma \approx A^{4/3}$.  The exact scaling
can be determined by a Glauber calculation \cite{usPRC,strikmanblack};
for the $\rho^0$, the scaling is roughly $A^{5/3}$.

The large photon flux and the coherent scattering lead to a large
cross section for $\rho^0$ production.  For 200 GeV per nucleon Au-Au
collisions at RHIC, the cross section for $\rho^0$ production is 590
mb, or about 8\% of the hadronic cross section \cite{usPRC}.  The STAR
collaboration has already published results on $\rho^0$ production in
Au-Au collisions at an energy of 130 GeV per nucleon \cite{STARrho}.
The cross sections agreed with the Glauber calculations.

The elastic scattering involves the strong force, and so has a very
short range \cite{muller}.  The $\rho$ production is localized to
within 1 fm of the two ions.  There are two possibilities: either
nucleus 1 emits a photon which scatters off nucleus 2, or vice versa.
These two possibilities are indistinguishable, and are related by a
parity tranformation.  Vector mesons have negative parity, so the two
amplitudes subtract, with a transverse momentum ($p_T$) dependent
phase factor to account for the separation.  The cross section is
\cite{interfere}
\begin{equation}
\sigma = \big|A_1 - A_2\exp{(ip_T\cdot b)}\big|^2
\end{equation}
where $A_1$ and $A_2$ are the amplitudes for $\rho^0$ production from
the two directions.  At mid-rapidity $A_1=A_2$ and this simplifies to
\begin{equation}
\sigma = \sigma_0 \big[(1-\cos{(p_T\cdot b)}\big]
\label{eq:interfere}
\end{equation}
where $\sigma_0$ is the cross section without interference.  The
system acts as a 2-slit interferometer, with slit separation $b$
\cite{prb}.  Of course, $b$ is unmeasurable, and the observed $p_T$
spectrum is obtained by integrating Eq. (1) over $b$.  The $p_T$
spectrum is suppressed for $p_T < \hbar/\langle b\rangle$, where
$\langle b\rangle$ is the median impact parameter.

The $\rho$ decay distance, $\gamma\beta c\tau < 1$ fm is far less than
$\langle b\rangle\approx 46$ fm.  The $\rho^0$ decay before the two
amplitudes can overlap, making this an interesting test of quantum
mechanics \cite{physlett}.

Two-photon processes like $e^+e^-$, Fig. \ref{diagram}b production are
purely electromagnetic.  To lowest order, it involves 2 photons, so
the cross section scales as $Z^4$.  At RHIC, the cross section is
expected to be huge - 33,000 barns for gold beams.  The coupling
constant $Z\alpha\approx 0.6$ is large, so perturbative descriptions
of this process may not be adequate.  The calculated Coulomb
corrections to the lowest order diagram reduce the cross section by
25\% \cite{ivanov}.  However, recent all-orders calculations find a
cross section that matches the lowest order result \cite{nonpert}.
The apparent discrepancy is not well understood \cite{problem}.

In UPCs, the coupling constant $Z\alpha$ (where $\alpha\approx 1/137$
is the electromagnetic coupling constant) is very large, so multiple
interactions between a single ion pair are common.  $\rho^0$ or
$e^+e^-$ production may be accompanied by mutual Coulomb excitation,
as in Figs. \ref{diagram}c and \ref{diagram}d.   Each ion emits a
photon, which excites the other nucleus, usually to a giant dipole
resonance (GDR).  The excited nucleus decays by emitting one or more
neutrons.  These neutrons serve as a distinctive experimental
signature of nuclear dissociation.  The presence of mutual Coulomb
dissociation can be used to `tag' collisions at moderate impact
parameters, $2R_A < b < \approx 30$ fm \cite{baltzus}.  Except for the
shared impact parameter, the reactions are independent
\cite{factorize}.  For example, the cross section for $\rho^0$
production with mutual Coulomb dissociation is
\begin{equation}
\sigma = \int d^2b P_\rho(b) P_{2EXC}(b) [1-P_{had}(b)]
\label{eq:factorize}
\end{equation}
where $P_\rho(b)$ and $P_{2EXC}(b)$ are the probabilities for $\rho^0$
production and mutual excitation respectively. Here, $P_{had}(b)$ is
the probability of having a hadronic interaction; the last term has an
effect similar to setting a minimum impact parameter $b_{min} = 2
R_A$.  For gold ions, $P_\rho(2R_A)\approx 1\%$ and
$P_{2EXC}(2R_A)\approx 30\%$, so the probability of multiple
interactions is substantial.  The multiple photon exchange skews these
interactions to smaller $b$.  This leads to a harder photon spectrum,
and as can be seen from Eq. (\ref{eq:interfere}), interference at
larger $p_T$.

\section{The STAR Detector}

The STAR detector studies heavy ion and polarized proton collisions at
the Relativistic Heavy Ion Collider (RHIC) at Brookhaven National
Laboratory.  STAR is optimized to study central heavy ion collisions,
which may contain thousands of particles \cite{STARoverview}.
However, it is also quite effective for studying UPCs where the final
states contain 2-4 particles.  STAR has collected UPC data in
gold-gold collisions at energies of 130 GeV per nucleon (in 2000) and
200 GeV per nucleon (in 2001).  STAR has also studied $\rho^0$
photoproduction in 200 GeV per nucleon deuteron on gold collisions
\cite{falk}.  In $dA$ collisions, the photon usually comes from the
gold nucleus, removing the photon direction ambiguity present in the
gold-gold collisions.  This writeup will focus on the 200 GeV per
nucleon gold-gold collisions.

Charged particles are reconstructed in a 4.2 meter long, 4 meter
diameter time projection chamber (TPC) \cite{TPC}.  The TPC is in a
solenoidal magnetic field which has been operated at both 0.25 and 0.5
T.  At 0.5 T, the reconstruction efficiency for charged pions is high
for tracks with transverse momentum $p_T>100$ MeV/c and pseudorapidity
$|\eta|<1.15$.  The track position and specific energy loss, $dE/dx$
were measured at 45 points for high momentum charged particles with
$|\eta|<1$.  In a 0.5 T field, the $dE/dx$ resolution was 8\%. The TPC
is surrounded by 240 scintillator slats covering $|\eta|<1$,
comprising the central trigger barrel (CTB).  Two zero degree
calorimeters (ZDCs) are located 18 m upstream and downstream of the
interaction point.  These calorimeters are sensitive to neutrons from
mutual Coulomb dissociation \cite{ZDCs}.

Two different triggers \cite{trigger} were used to study UPCs.  The
`topology' trigger selected events with an appropriate topology in the
CTB.  It divided the CTB into 4 quadrants: north, south, top and
bottom, and required hits in the north and south quadrants.  The top
and bottom regions were used as vetoes to reject cosmic rays.

The minimum bias trigger selected events with one or more neutrons in
each ZDC.  The ZDC signals were required to occur within 1 nsec,
restricting these events to the central 30 cm of the TPC.  Data from
both triggers was processed identically, except that events from the
CTB based trigger were distributed more broadly along the TPC axis,
and consequently, were accepted in a broader range.

\section{$\rho^0$ Production}

UPC $\rho^0$ production has a distinctive signature - two oppositely
charged tracks with small net $p_T$.  We select events with exactly 2
primary tracks that form a vertex.  The vertexing procedure considers
all of the tracks in an event, and rejects tracks that are
inconsistent with coming from a single vertex.  We do allow a few
non-primary background tracks in the TPC. The topology and minimum
bias data were treated identically, except that allowance was made for
the different distribution of the accepted event production
points. This study used about 1.5 million minimum bias and about 1.7
million topology triggers taken in a 0.5 T field for 200 GeV per
nucleon interactions.

Figure 2a shows the $p_T$ distribution of all charge-0 2-track
vertices.  The large peak for $p_T < 100$ MeV/c is a signature of
fully coherent interactions. Like-sign pion pairs (the shaded
histogram) are used as a background estimate; they are normalized to
fit the unlike-sign pairs for $p_T > 250$ MeV/c.  This procedure
treats incoherent $\rho^0$ production (a photon scattering from a
single nucleon in the target) as background.

Figure 2b shows the rapidity distribution of the coherent $\rho^0$
(pairs with $p_T < 150$ MeV/c).  The points (data) are in excellent
agreement with a calculation based on the soft Pomeron model and our
detector simulation.

Figure 2c shows the $\pi\pi$ invariant mass, $M_{\pi\pi}$ for the
sample.  The data is fit to a relativistic Breit-Wigner for the
$\rho^0$, plus a S\"oding term to account for the interference with
direct $\pi^+\pi^-$ production\cite{soding}.  The interference shifts
the peak of the distribution to lower $M_{\pi\pi}$. The direct
$\pi\pi$ to $\rho$ ratio agrees with the STAR 130 GeV analysis
\cite{STARrho} and with that observed by the ZEUS collaboration in
$\gamma p$ interactions \cite{ZEUS}.  However, because of the coherent
enhancement, the STAR data is at smaller $|t|$ than the ZEUS analysis,
so the two results may not be directly comparable \cite{marks}.

\begin{figure}
\label{rhoplots}
\setlength{\epsfxsize=2.9 in} 
\setlength{\epsfysize=2.1 in} 
\centerline{\epsffile{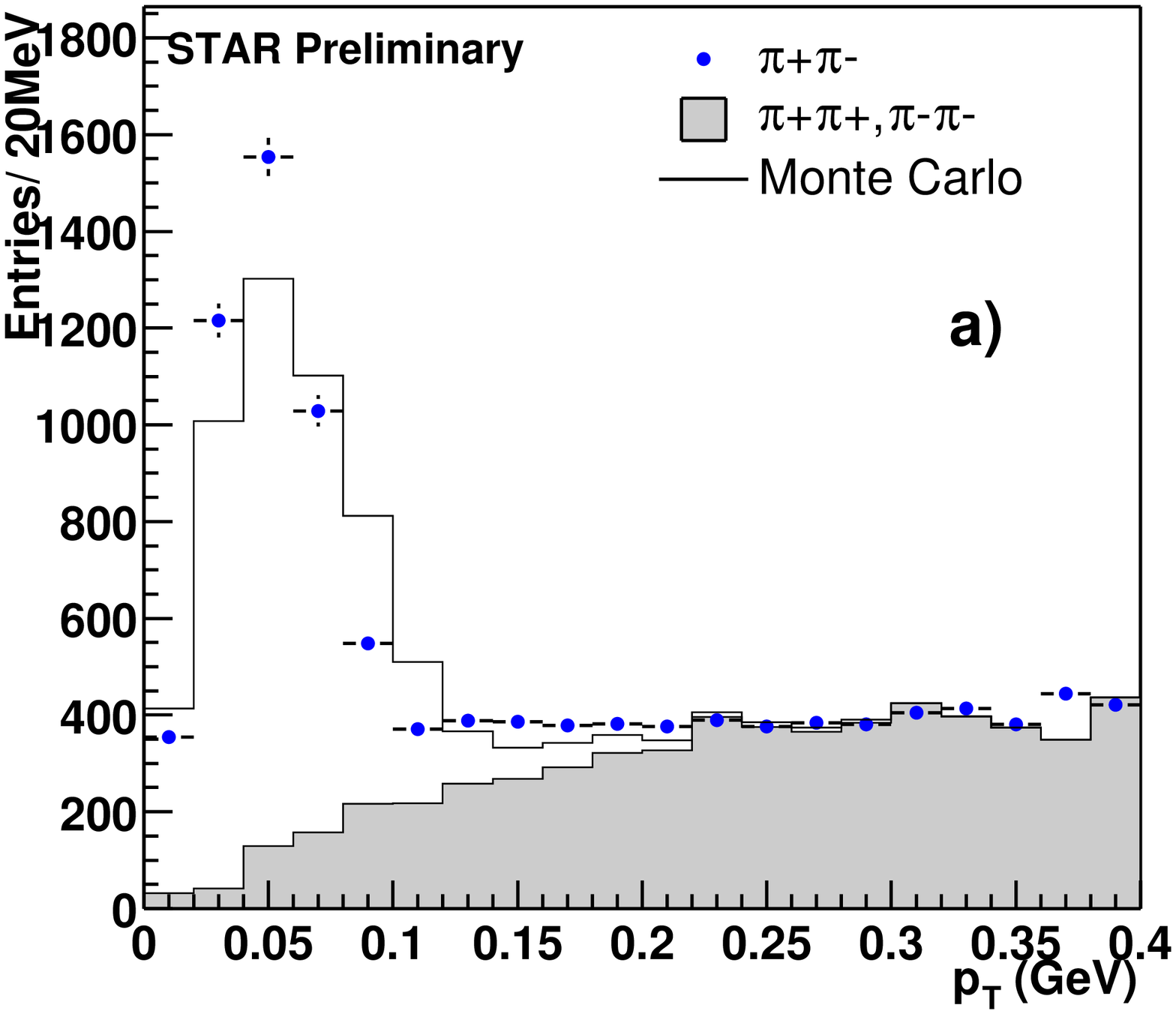}}
\centerline{\epsffile{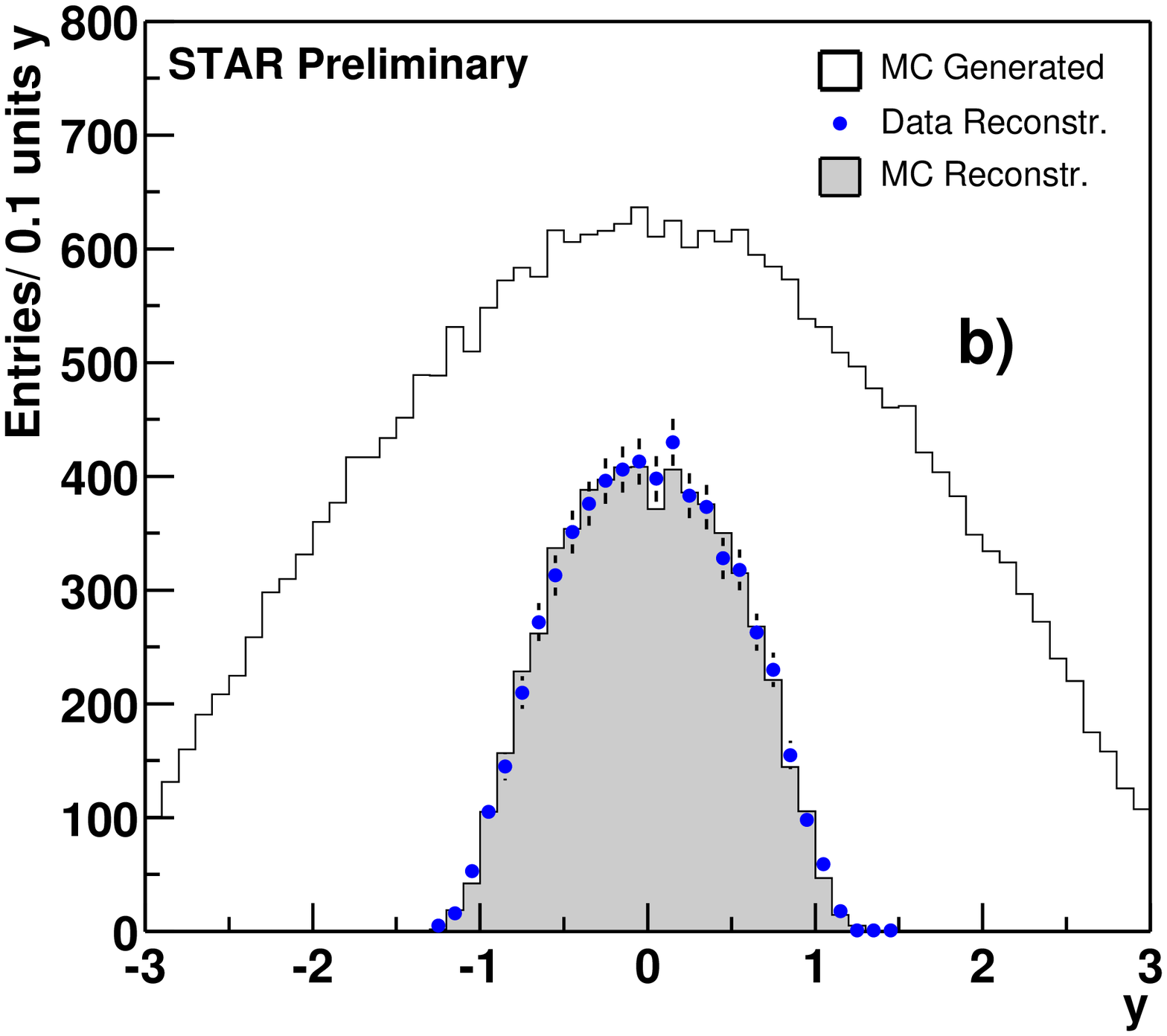}}
\centerline{\epsffile{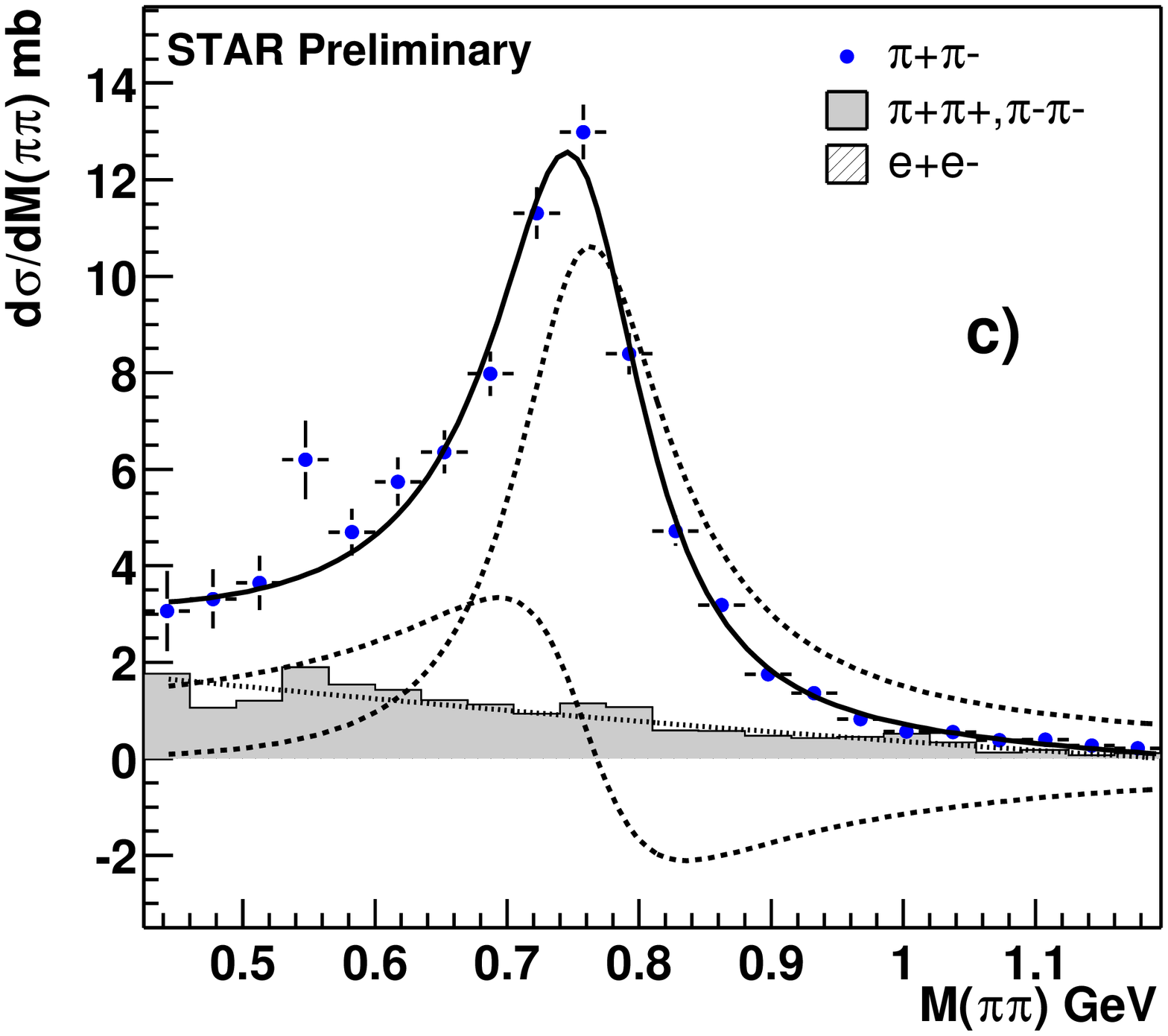}}
\caption[]{The (a) $p_T$, (b) rapidity and (c) $\pi\pi$ invariant mass
spectra of $\rho^0$ using data from the minimum bias trigger.  The
mass is fit to to a Breit-Wigner function for the $\rho^0$, plus a
{S\"oding} term due to interference with a direct $\pi\pi$ component.
The combinatoric hadronic and coherent $e^+e^-$ pair backgrounds are
also shown.
}
\end{figure}

\section{$\rho^0$ Interferometry}

For the interference analysis, a clean, low-background sample is
important, so stringent event selection criteria were used, at some
cost in efficiency.  Events were required to have exactly two tracks
with a vertex within 50 cm longitudinally of the center of the TPC for
the minimum bias sample, and 100 cm for the topology sample.  $\rho^0$
candidates were further required to have rapidity $|y|>0.1$, to
eliminate a small remaining cosmic ray contamination.  Finally, the
pairs were required to have 550 MeV $< M_{\pi\pi} < 920$ MeV.  The
minimum mass cut removed most of the hadronic backgrounds, which are
concentrated at low $M_{\pi\pi}$.  These cuts reduced the like-sign
pair background to a few percent. With these cuts, background from
misidentified two-photon production of lepton pairs should be very
small.  The sample still includes direct pions, which should have the
same spin/parity and quantum mechanical behavior as the pion pairs
from $\rho$ decay.  We do not distinguish between the two sources.

The interference depends on the amplitudes for $\rho$ production on
the two nuclei.  Away from $y=0$, the photon energies for the two
photon directions differ, $k_{1,2} = M_V/2 \exp(\pm y/2)$; the
amplitudes differ and the interference is less than maximal.  Although
it is not expected in the soft-Pomeron model, the the photon energy
difference could introduce a small $\rho^0$ production phase
difference, which could affect the interference \cite{phase}.  This
analysis focuses on the region near mid-rapidity where any phase
difference is small.  A Monte Carlo calculation is used to find the
interference for different rapidity ranges \cite{usPRC,interfere}.

We use the variable $t_\perp = p_T^2$ to study the interference.  At
RHIC energies, the longitudinal component of the 4-momentum transfer
is small, so $t\approx t_\perp$.  Without interference, the spectrum
$dN/dt \approx\exp{(-bt)}$ for a variety of nuclear models
\cite{interfere,ting}.  Our calculations use a Woods-Saxon
distribution for the gold density distribution.

Figure 3 compares the uncorrected minimum bias data for $0.1 <
|\eta|<0.5$ with two simulations, with and without interference.  Both
simulations include the detector response.  The data has a significant
downturn for $t<0.001$ GeV$^2$, consistent with the $\langle b\rangle
= 20$ fm expected for a 3-photon reaction\cite{factorize}. This drop
matches the drop seen in the calculation with interference, but is
absent in the calculation without the interference.

\epsfclipon
\begin{figure}
\label{rawinterfere}
\setlength{\epsfxsize=3.0 in} 
\setlength{\epsfysize=2.1 in} 
\centerline{\epsffile{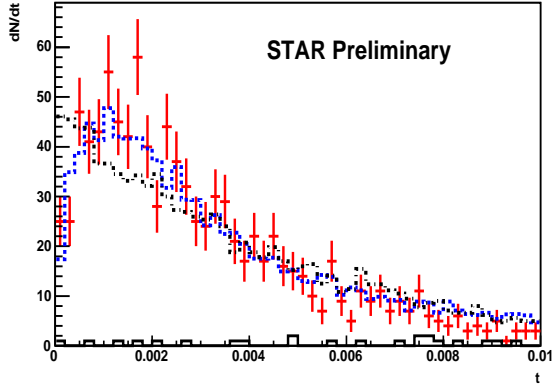}}
\caption[]{Raw (uncorrected) $t_\perp$ spectrum for $\rho^0$ sample
for $0.1 < |y|<0.5$ for the topology data.  The points (red) are the
data.  The dashed histogram (blue) is a simulation that assumes that
there is interference, while the dot-dash histogram (black) is based
on a calculation without interference.  The solid black histogram with
very few counts is the like-sign background.}
\end{figure}

The efficiency corrected data are shown in Fig. 4. Minimum bias and
topology data are shown separately, each with two rapidity bins: $0.1
< |y| < 0.5$ and $0.5 < |y| < 1.0$.  The efficiency is independent of
$p_T$.  However $p_T$ smearing (resolution) does affect the spectrum
slightly.  The $\rho^0$ $p_T$ resolution is about 9 MeV/c, while the
1st $t$ bin covers 0 to (15 MeV/c)$^2$.  Interference depletes the
first few bins, but feed down from the higher t bins partially
repopulates them.

\begin{figure*}
\label{tspectra}
\setlength{\epsfxsize= 2.9 in} 
\setlength{\epsfysize=2.1 in}
\centerline{\epsffile{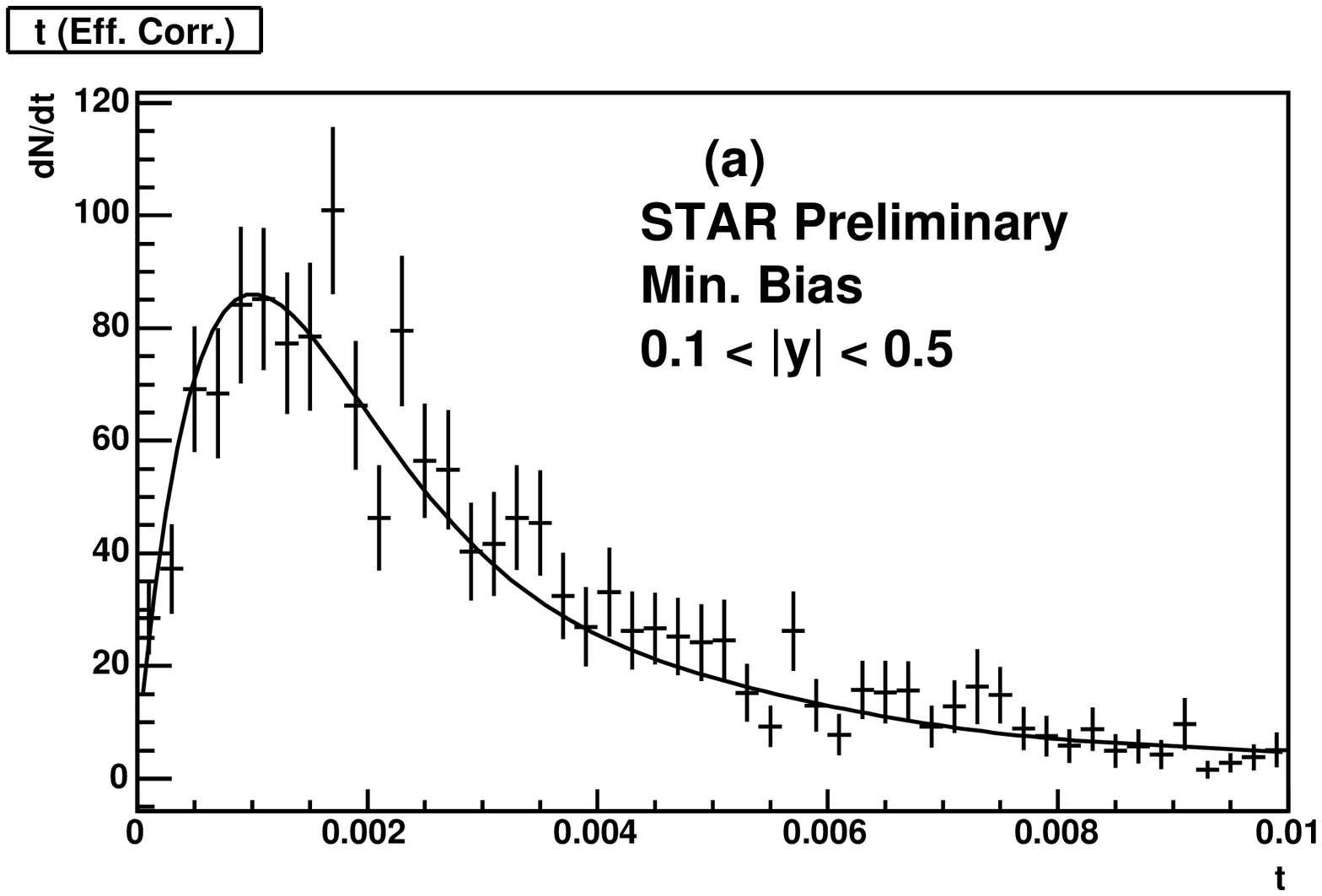}
\setlength{\epsfxsize= 2.9 in} 
\setlength{\epsfysize=2.1 in}
\epsffile{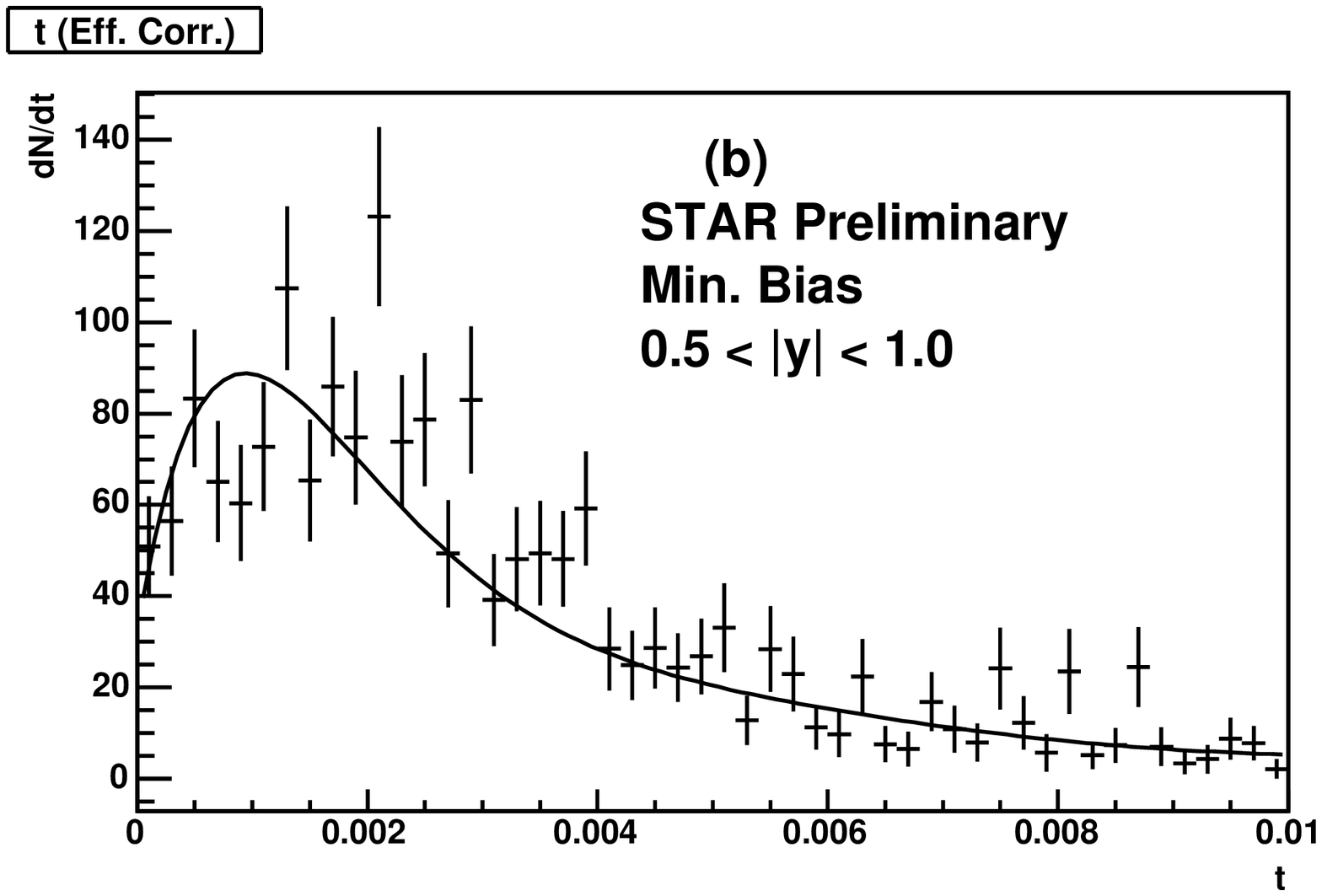}}
\vskip .01 in
\setlength{\epsfxsize= 2.9 in} 
\setlength{\epsfysize=2.1 in}
\centerline{\epsffile{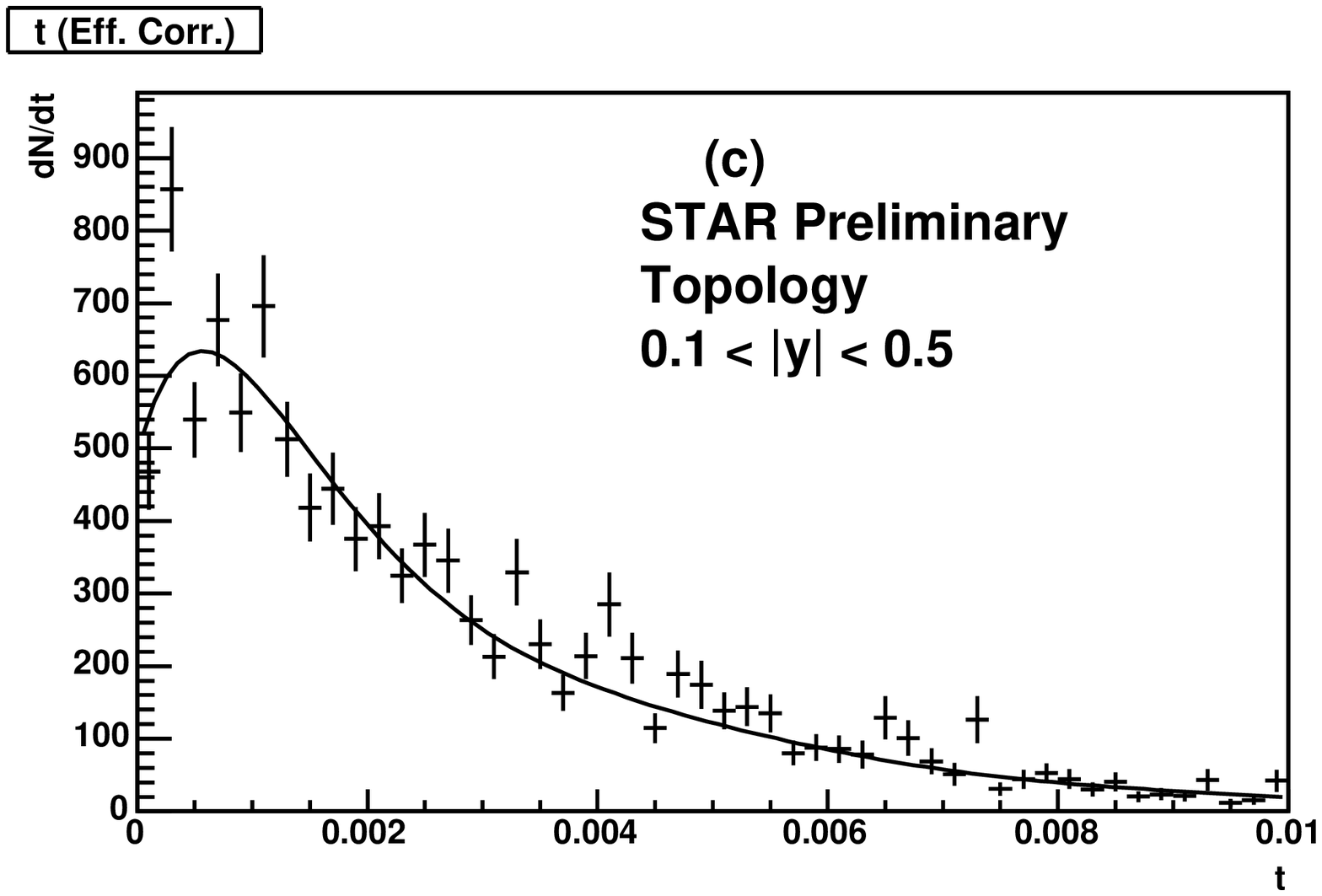}
\setlength{\epsfxsize= 2.9 in} 
\setlength{\epsfysize=2.1 in}
\epsffile{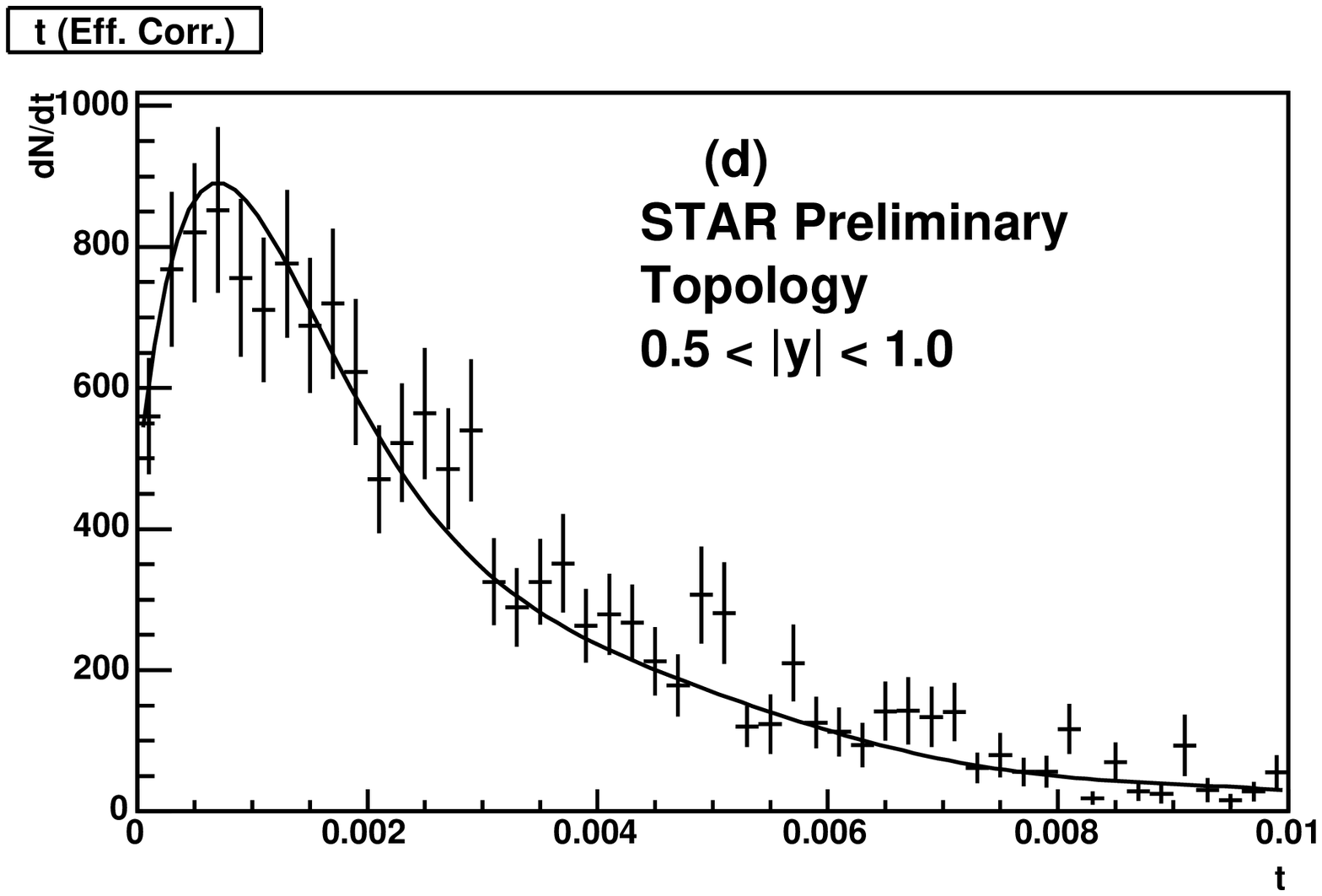}}
\caption[]{Efficiency corrected $t_\perp$ spectrum for $\rho^0$ from
(a) mutual dissociation with $0.1 < |y|<0.5$, (b) mutual dissociation
with $0.5 < |y|<1.0$, (c) topology trigger with $0.1 < |y|<0.5$ and
(d) topology trigger with $0.5 < |y|<1.0$.  The points are the data,
while the solid line is the result of the fit to Eq. 4.}
\end{figure*}

The data is fit to the 3-parameter form:
\begin{equation}
{dN\over dt} = a \exp(-bt) [1+ c(R(t)-1]
\end{equation}
where $R(t) = Int(t)/Noint(t)$ is the ratio of the Monte Carlo
t-spectra with and without interference.  Here, $a$ is the overall
normalization, the slope $b\approx R_A^2$, and $c$ is the degree of
spectral modification; $c=0$ corresponds to no interference while
$c=1$ is the expected interference.  This functional form separates
the interference ($c$) from the nuclear form factor ($b$).

Table 1 gives the results of the fits.  At small-rapidities the
amplitudes from the two directions are similar and the interference
reduces the cross section at $p_T=0$ by more than at larger
rapidities.  In the minimum bias data, the interference extends to
higher $p_T$ than the topology data because the former has a smaller
$\langle b\rangle$.

\begin{table}
\begin{tabular}{lcrr}
Trigger & Rapidity & \multicolumn{1}{c} b (GeV$^{-2}$) & \multicolumn{1}{c}c \\
\hline
M.B. & $0.1 - 0.5$&  $301\pm14$  & $1.01\pm 0.08$ \\
M.B. & $0.5 - 1.0$&  $304\pm 15$ & $0.78\pm 0.13$ \\
Topo. & $0.1 - 0.5$&  $361\pm 10$ & $0.71\pm0.16$ \\
Topo. & $0.5 - 1.0$&  $368\pm 12$ & $1.22\pm0.21$ \\
\hline
\end{tabular}
\caption[]{Results of the fits to the minimum bias (M.B.) and topology
(topo.) data for two rapidity bins.  The fits have $\chi^2$ of 50, 73,
81 and 50 respectively for 47 degrees of freedom.}
\label{t:fitout}
\end{table}

The 4 $c$ values are consistent within errors; the weighted
average is $c = 0.93 \pm 0.06$.  The $b$ values for the
minimum bias and exclusive $\rho^0$ data differ by 20\%: $364\pm7$
GeV$^{-2}$ for the exclusive $\rho$ versus $303\pm 10$ GeV$^{-2}$
for the Coulomb breakup events.  

The different $b$ values may be attributed to the different impact
parameter distributions caused by the nuclear breakup tagging in the
minimum bias data.  The photon flux depends on the impact parameter as
$1/b^2$.  When $b\approx {\rm few} R_A$, $\rho$ are more likely to be
produced on the side of the target near the photon emitter than on the
far side.  $\rho^0$ production is concentrated on the near side,
leading to a smaller effective production volume and the smaller $b$.
This near-side skewing affects the interference slightly, but is not
included in current calculations.

Systematic errors come from a variety of sources.  We have studied the
effect of detector distortions (primarily $p_T$ smearing) by turning
off the detector simulation and comparing raw simulations with
reconstructed data.  This lowered $c$ by 18\%; if the detector
simulation is 75\% correct (a very conservative assumption), then
detector effects are a less than 5\% systematic uncertainty. Minor
systematic uncertainties come from backgrounds and the fitting
procedure.  We have compared the data and simulations for a variety of
kinematic and detector-based variables, and found good agreement. We
estimate a preliminary experimental systematic uncertainty of 8\%.

This analysis depends on the shape of $R(t)$, which is calculated
following Ref. \cite{interfere}.  For the broader picture, it is
necessary to consider the simplifications in Ref. \cite{interfere}.
The calculation averages the photon flux over the surface of the
nucleus, rather than incorporating the proper $1/b^2$ weighting.  This
may partly explain the relatively poor fit for the large-$|y|$ Coulomb
breakup fit.  We estimate that the effective impact parameter should
be within 10\% of the actual ion-ion separation for the coulomb
breakup data, and 3\% for the exclusive $\rho$.  The uncertainties in
the calculations should be at most a 15\% effect. With this, the
interference is $93\pm6 (stat.) \pm 8 (syst.) \pm 15 (theory) \%$
(STAR preliminary) of that expected.

The $\rho$ decays rapidly, with $\gamma\beta c\tau\ll \langle
b\rangle$ and the two $\rho$ decay points are well separated in
space-time.  In the usual space-time picture, the decays occur
independently, and any interference must involve the final state
$\pi^+\pi^-$.  Interference must involve identical final states from
the two sources.  However, given the large available phase space for
the decays, this is very unlikely for independent decays.

One possible interpretation of this result is given in
Ref. \cite{physlett}. In it, the interference occurs because the
post-decay wave function includes amplitudes for all possible final
states, then the amplitudes for identical states subtract, and the
interference is visible.  Because of the two sources, the $\pi^+\pi^-$
wave function is non-factorizable, and thus exhibits the
Einstein-Podolsky-Rosen paradox \cite{EPR}.  We find that the
decoherence, $1-c$, due to environmental or other factors is less than
43\% at the 90\% confidence level.

\section{$e^+e^-$ Pairs}

The cross section for the production of $e^+e^-$ pairs is peaked near
threshold, with pair mass $M_{ee}\approx 3.5 m$, where $m$ is the
electron mass.  The pairs are produced predominantly in a
forward-backward topology, with the $e^\pm$ produced with large
longitudinal (along the beampipe) momentum and small $p_T$.  This
complicates experimental detection at an ion collider, and reduces the
statistics available to any analysis.  This analysis used data about
800,000 minimum bias triggers taken in a 0.25 T magnetic field.  The
lower magnetic field (compared to the 0.5 T full field data) increased
the sensitivity to low $p_T$ particles. The minimum bias trigger freed
the electrons from the requirement that they have a high enough $p_T$
to reach the CTB.

This analysis \cite{vladimir} used tracks with $p_T > 65$ MeV/c and
pseudorapidity $|\eta| < 1.15$.  In this region, tracking efficiency
for electrons was above 80\%.  The electrons were identified by
$dE/dx$.  To get good electron/hadron separation, we required the
tracks to have $p<130$ MeV/c.  In this momentum region, electrons had
considerably higher $dE/dx$ than hadrons, and the identification
efficiency was nearly 100\%, with minimal contamination.  The cross
section falls steeply with increasing $M_{ee}$, so few leptons were
expected with higher momenta.  We also required the events to have
pair $p_T < 100$ MeV/c, pair rapidity $|y|<1.15$ and pair mass 140 MeV
$< M_{ee} < 265$ MeV.  These cuts selected a sample of 52 events.

\begin{figure*}
\label{pairs}
\setlength{\epsfxsize=2.9 in} 
\setlength{\epsfysize=2.1 in}
\centerline{\epsffile{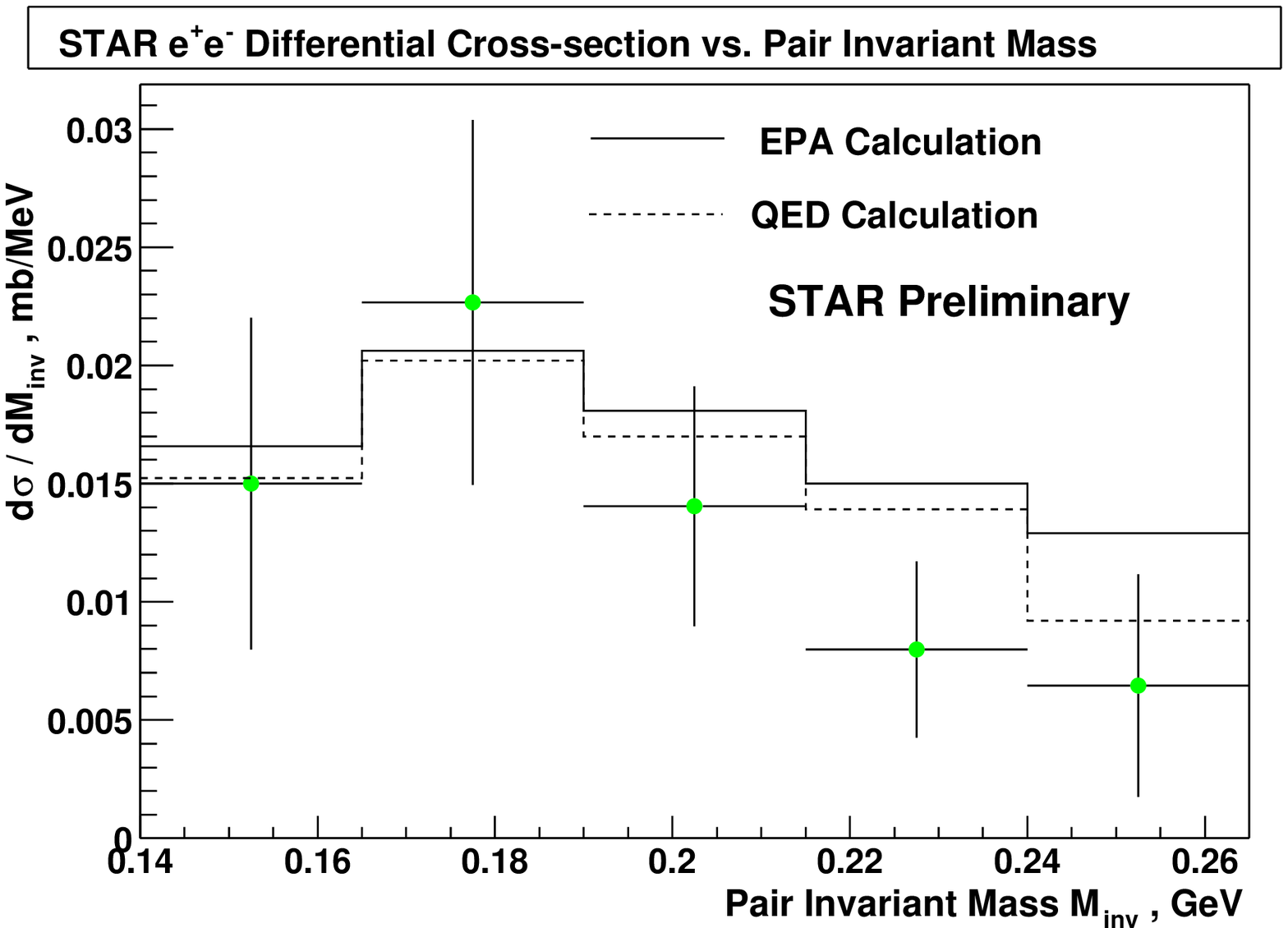}
\setlength{\epsfysize=2.1 in}
\setlength{\epsfxsize=2.9 in} 
\epsffile{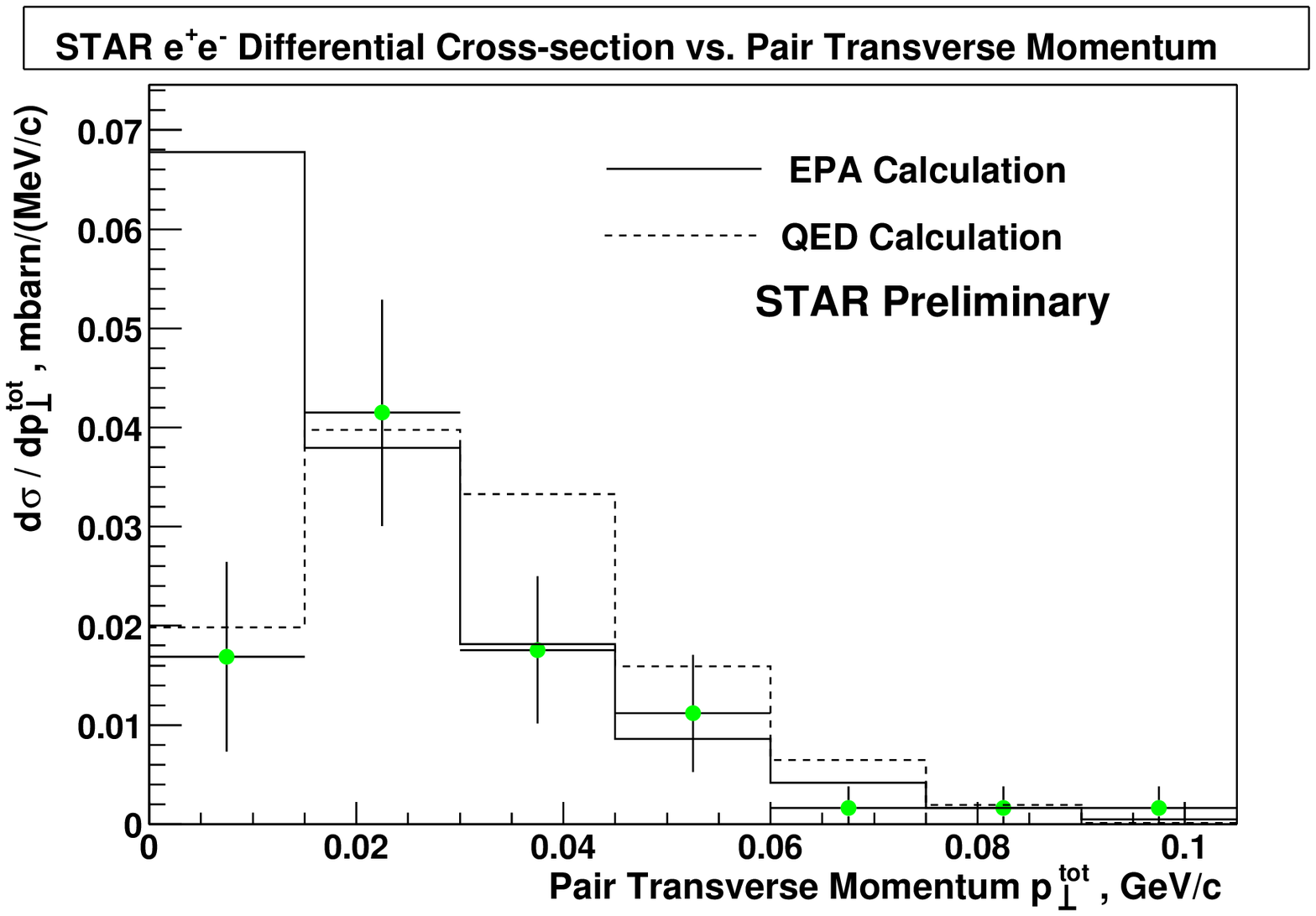}
}
\caption[]{Differential Cross sections for pair (a) mass and (b)
$p_T$.  The data (points) are compared with the results of a
equivalent photon calculation (solid) and a quantum electrodynamics
calculation (dashed line).  All data and calculations are restricted
to the kinematic region described in the text.}
\end{figure*}

The major backgrounds in this analysis are due to misidentified
coherent $\pi^+\pi^-$ pairs and incoherent hadronic backgrounds.  The
former can be calculated from the known cross section and
misidentification probability. The latter can be studied by using like
sign ($\pi^+\pi^+$) pairs.  The total background is estimated to be
about 1.1 events.

The integrated luminosity of $94\pm 9$ mb$^{-1}$ was estimated by
counting hadronic events with at least 8 tracks.  This selected 
80\% of all hadronic gold-gold interactions.  The integrated
luminosity is based on an assumed gold-gold hadronic cross section
of 7.2 barns.

Figure 5 shows the cross section as functions of pair mass and $p_T$.
The data is compared with two calculations.  The first, shown by the
solid line, uses Eq. 3, with a equivalent photon calculation for
$P_{ee}(b)$.  The second, provided by Kai Hencken, was a quantum
electrodynamic calculation, including photon virtuality. The major
difference between the results of the two calculations occurs for $p_T
< 15$ MeV/c.  The data strongly favors the QED calculation, by
$4\sigma$.  

As expected, the $\langle p_T \rangle$ for $e^+e^-$ production is 
smaller than for $\rho^0$ production.  This is because the Pomeron
$p_T$ scale is roughly $\hbar/R_A$, while the photon $p_T$ scale
is very roughly $M_{ee}/\gamma\approx \hbar/b$. 

Within the kinematic acceptance region, we find a cross section
$\sigma = 1.6 \pm 0.2 \pm 0.3$ mb (STAR preliminary).  The major
systematic errors are due to tracking and vertexing uncertainty, and
the integrated luminosity.  This is 1.2$\sigma$ lower than the
equivalent photon prediction of 2.1 mb, and $0.8\sigma$ below the QED
calculation of $\sigma_{QED} = 1.9$ mb for the same kinematic cuts.
This measurement and the QED calculation can be used to put limits on
possible higher order corrections $\Delta\sigma$ to the cross section.
At a 90\% confidence level, we find $-0.5\sigma_{QED} < \Delta\sigma <
0.2\sigma_{QED}$.

\section{Conclusions}

STAR has studied a number of photonuclear and two-photon reactions.
We observe both exclusive $\rho^0$ production and $\rho^0$ production
accompanied by nuclear excitation.  The cross sections and rapidity
distributions match the predictions of the soft Pomeron model.  We
observe destructive interference between the two production sites, at
the expected level, and set an upper limit on decoherence.  Finally,
we observe $e^+e^-$ pair production accompanied by mutual Coulomb
dissociation.  The kinematic distributions match those predicted by
lowest order quantum electrodynamics.

We thank Tony Baltz for providing the calculations of nuclear
breakup and Kai Hencken for providing the QED calculations.  
This work was supported by the U.S. DOE under contract number 
DE-AC-03-76SF00098.


\begin{thebibliography}{99}
\def\etal{{\it et al.}}

\bibitem{reviews}G. Baur {\it et al.}, Phys. Rep. {\bf 364}, 359
(2002); F. Krauss, M. Greiner and G. Soff, Prog. Part. Nucl.
Phys. {\bf 39}, 503 (1997).

\bibitem{phase}T. H. Baur {\it et al.}, Rev. Mod. Phys.  {\bf 50}, 261
(1978).

\bibitem{strikman}L. Frankfurt, M. Strikman and M. Zhalov,
Phys. Lett. {\bf B540}, 220 (2002); L. Frankfurt {\it et al.}, JHEP
{\bf 8}, 43 (2003).

\bibitem{usPRC}S. Klein and J. Nystrand, Phys. Rev. {\bf C60}, 014903
(1999).

\bibitem{strikmanblack} L. Frankfurt, M. Strikman and M. Zhalov,
Phys. Rev. {\bf C67}, 034901 (2003).

\bibitem{STARrho}C. Adler {\it et al.}, Phys. Rev. Lett. {\bf 89},
272302 (2002).

\bibitem{muller}B. M\"uller and A. J. Schramm, Nuclear
Physics {\bf A523}, 677 (1991).

\bibitem{interfere}S. Klein and J. Nystrand, Phys. Rev. Lett.
{\bf 84}, 2330 (2000).

\bibitem{prb} An analagous two-slit interferometer is described by
T. Sudbery in {\it Quantum Concepts in Space and Time}, ed. R. Penrose
and C. J. Isham, (Oxford, 1986).

\bibitem{physlett}S. Klein and J. Nystrand, Phys. Lett. {\bf A308},
323 (2003).  

\bibitem{ivanov}D. Yu Ivanov, A. Schiller and V. Serbo,
Phys. Lett. {\bf B454}, 155 (1999).

\bibitem{nonpert}A. J. Baltz and L. D. McLerran, Phys. Rev.
{\bf C58}, 1679 (1998).

\bibitem{problem}A. Aste {\it et al.}, Eur. Phys. J.  {\bf C23}, 545
(2002).

\bibitem{baltzus}A. Baltz, S. Klein and J. Nystrand,
Phys. Rev. Lett. {\bf 89}, 012301 (2002).

\bibitem{factorize}G. Baur {\it et al.}, nucl-th/0307031.

\bibitem{STARoverview}K. H. Ackermann {\it et al.}, Nucl. Instrum.
\& Meth. {\bf A499}, 624 (2003).

\bibitem{falk}F. Meissner and V. B. Morozov, nucl-ex/0307006 (2003).

\bibitem{TPC}M. Anderson \etal, Nucl. Instrum. \& Meth. {\bf A499}, 659
(2003); M. Anderson \etal, Nucl. Instrum. \& Meth. {\bf A499}, 679
(2003).

\bibitem{ZDCs}C. Adler {\it et al.}, Nucl. Instrum. \& Meth.  {\bf
A470}, 488 (2001).

\bibitem{trigger}F. S. Bieser {\it et al.}, Nucl. Instrum. \&
Meth. {\bf A499}, 766 (2003).

\bibitem{soding}P. S\"oding, Phys. Lett. {\bf 19}, 702 (1966).

\bibitem{ZEUS}J. Breitweg {\it et al.}, Eur. Phys. J.
{\bf C15}, 1 (2000).

\bibitem{marks}Mark Strikman, private communication (2003).

\bibitem{ting}M. Alvensleben {\it et al.}, Phys. Rev. Lett.
{\bf 24}, 792 (1970).

\bibitem{EPR}A. Einstein, B. Podolsky and N. Rosen, Phys. Rev.
{\bf 47}, 777 (1935).

\bibitem{vladimir}V. Morozov, PhD Dissertation, UC Berkeley, 2003.

\end{thebibliography}
\end{document}